\documentclass[twocolumn,twoside,letterpaper,11pt]{article}
\usepackage{graphicx,xrrc,natbib}
\usepackage{times}

\def\la{\mathrel{\hbox{\rlap{\hbox{\lower4pt\hbox{$\sim$}}}\hbox{$<$}}}}
\def\ga{\mathrel{\hbox{\rlap{\hbox{\lower4pt\hbox{$\sim$}}}\hbox{$>$}}}}

\begin{document}

\title{OUTSTANDING QUESTIONS ABOUT MERGERS IN CLUSTERS OF GALAXIES}

\author{    C. L. Sarazin                                } 
\institute{ University of Virginia                       } 
\address{   P.O. Box 3818, Charlottesville, VA 22903-0818, U.S.A.        } 
\email{     sarazin@virginia.edu }

\maketitle

\abstract{At the request of the organizers of the Santa Fe meeting on
X-Ray and Radio Connections, I list some important questions and lines
of inquiry for the future study of the X-ray and radio connections
in mergers in clusters of galaxies, These questions are based on the
talks, posters, and discussions at this meeting,}

\section*{Questions}
\label{sec:questions_sarazin_intro}

\begin{enumerate}

\item {\bf Can we construct a rigorous quantitative definition of a
``merging cluster''?}
As the quality of the X-ray data improves, we need a better definition
since most clusters show some evidence for the effects of mergers.
It is unlikely that a simple bifurcation between merging and non-merging
clusters is useful.
We need a more continuous, quantitative measure or measures of the
degree to which mergers affect the dynamics in each cluster.
One simple example are the ``power-ratios'' developed by Buote.

\item {\bf How is shock acceleration affected by the low Mach numbers in
merger shocks?}  Is the primary effect just the lower compression and
steeper particle spectra predicted by kinematic theory?  Or, is there
a critical Mach number below which shock acceleration of electrons
is doesn't occur?

\item {\bf What is the source of the seed electrons for shock or
turbulent acceleration in clusters?}  Are they thermal, or
low energy relativistic electrons from radio sources or previous mergers?

\item {\bf Are cluster radio relic sources due to (re)acceleration
of electrons by merger shocks?
Or, are they produced by the compression of old radio sources
(``radio ghosts'')?}
How can we determine which is correct?
One way would be to compare radio and X-ray images  and spectra
in detail.
The X-ray images could show us the locations of the shocks.
If the relics are due to shock acceleration, they should have sharper
edges and flatter spectra on the shock side.
Better radio images and spectral index maps are needed.
In the X-ray, it is difficult to study the regions around relics as
they are often in the outer parts of clusters where the X-ray surface
brightness is very low.

\item {\bf Are cluster radio halos due to turbulent acceleration?
Since they are only present in merging clusters, does this mean that
non-merging clusters have quiescent intracluster gas?}
We can test this with ASTRO-E2 and other observations of turbulent
motions in the the ICM.

\item {\bf Why are radio halos and relics comparatively rare?}
Are they really only found in merging clusters?
Given the strong correlation between radio halo power and
$L_X$ or $T_X$ for merging clusters, {\bf is there a large population
of fainter radio halos and relics?}
Is so, they may be found with more sensitive searches with LOFAR,
LWA, EVLA, and other instruments.

\item {\bf Do all merging clusters have radio halos and/or relics?}

\item {\bf How does fossil radio plasma from old radio sources get
distributed in clusters?}
Is it lifted buoyantly by radio bubbles?
Does the fossil radio plasma mix with the thermal ICM, or does in
remain in small or large discreet regions?

\item {\bf How turbulent is the intracluster medium?  Are merging clusters
much more turbulent?}
Hopefully, we will be able to measure turbulent velocities in the X-ray
gas with ASTRO-E2.

\item {\bf Which regions of clusters are turbulent?}
Observations of cold fronts suggest the gas is laminar except in the
wake.

\item Although the energy density in turbulence in clusters is
probably smaller than the thermal energy density, {\bf can turbulence
heat the thermal ICM in some locations?}
Is this heating important in cooling flows?

\item {\bf What are the radio mini-halos in the centers of some clusters?}
Are they related to cluster radio halos?
Are they produced by particle acceleration in cooling flows, where
they are often found?
Or, are they simply due to some sort of leakage of relativistic electrons
from the central radio sources in these cooling flows?

\item {\bf What is the origin of the ``sloshing features'' seen in Chandra
images of the central regions of many (particularly cooling flow) clusters
of galaxies?  Are they related to mergers?}
Hydrodynamical simulations would be very useful to determine the origin
and evolution of these sloshing features.

\item {\bf A library of mergers from hydrodynamical simulations would be
very useful.}
Fairly complete suites of merger simulations, perhaps extracted from
cosmological large scale structure simulation, are needed to study
clusters.

\item Simple analytical techniques have been used to estimate kinematic
parameters in simple merging clusters from X-ray observations.
It is clear that more detailed hydrodynamical simulations are needed to
study most cluster mergers.
{\bf Is there a simple way to find the relevant hydrodynamical model for
an observed cluster?}
We need an efficient technique to match simulations with
the observations of a given cluster.

\item {\bf Can we perform numerical simulations of cluster mergers with
much better resolution and with more of the relevant physics (magnetic
fields, cooling, turbulence, nonthermal particles, \dots )?}

\item {\bf Are the nonthermal particles in clusters dynamically important?}
At present, I would say the answer is ``probably not,'' but we won't
really know until we can detect the relativistic protons in clusters.

\item {\bf Are clusters of galaxies generally luminous gamma-ray sources?}
They are predicted to be by models for the relativistic particle
populations in clusters.
Will they be detected with AGILE and GLAST?
Can TeV gamma-rays be convincingly be detected with ground-based
instruments like VERITAS?
Will only merging clusters be detected (either at GeV or TeV energies)?

\item {\bf What is the nature of the EUV/soft X-ray excesses observed
towards clusters?}
Are they real, and how commonly do they occur?
Do they only occur in merging clusters?
What is the spatial distribution of the soft excess?
{\bf In general, is the emission thermal or nonthermal?}

\item {\bf What is the nature of the hard X-ray (HXR) excesses seen towards
a few clusters with BeppoSAX and RXTE?}
Are they real, and how commonly do they occur?
Do they only occur in merging clusters?
What is the spatial distribution of the soft excess?
Better information of the hard X-ray excesses may come soon from observations
with INTEGRAL, or eventually with future hard X-ray space missions.
{\bf Can we detect HXR excesses more easily in groups, where the
thermal X-ray emission is softer?}

\item {\bf What are the magnetic fields in clusters like?}
Can we reconcile the low values of $B$ found from the HXR with the
larger values based on Faraday rotation?
What is the spatial coherence length of the field?
Is it really possible to characterize the magnetic field in any regions
of a cluster by a typical value, or is there a (turbulent) spectrum
with many values of $B$ on different spatial scales?
How do the properties of the field vary with radius in a cluster?
How do magnetic fields vary with the merger state of the cluster?

\item {\bf What is the source of cluster magnetic fields?}
Are they generated from weak seed fields during cluster mergers
and formation?
Do they come from galaxies (probably not)?
Are they amplified or produced by plasma instabilities?

\item {\bf How do mergers disrupt cooling flows in cluster cores?}
What is the detailed physical mechanism (or mechanisms)?
How large a merger is required?
How long does the disruption last?

\item {\bf What, if anything, is the connection between mergers and
the life cycles of cluster-core radio sources?}

\item {\bf How do mergers affect the Sunyaev-Zeldovich (SZ) effect?}
How are cosmological conclusions based on SZ measurements affected?

\item {\bf Can we use SZ images to study outer merger shocks and accretion
shocks in clusters, and determine the pressure variations due to mergers?}
Since the SZ effect depends on pressure (and thus only linearly on density), in
principle SZ measurements could be more sensitive probes of
shocks in the outer parts of clusters, particularly in regions where
radio relics are found.

\item {\bf Do mergers significantly increase the cross-section for strong
lensing in clusters?}

\item {\bf What are the correct rates for transport processes in
clusters?}
What is the correct value of the thermal conductivity?
What is the correct value of the viscosity?
How are transport processes affected by magnetic fields?

\item {\bf Are clusters the (or an important or a) source of very
high energy cosmic rays?}

\item {\bf Can we extend the study of the thermal and nonthermal processes
in clusters of galaxies to supercluster, filaments, and also to groups
of galaxies?}

\end{enumerate}

\section*{Acknowledgments}
I thank Tracy Clarke and Maxim Markevitch for useful comments.
This work was supported by the National
Aeronautics and Space Administration through
Chandra Award Numbers
GO2-3159X,
GO3-4155X,
GO3-4160X,
GO4-5149X,
and
GO4-5150X,
and
XMM/Newton Award Numbers
NAG5-13737
and
NAG5-13088.

\end{document}